\documentclass{pr-imfp09}

\newcommand{\lsim}{\mathrel{\mathop{\kern 0pt \rlap
  {\raise.2ex\hbox{$<$}}}
  \lower.9ex\hbox{\kern-.190em $\sim$}}}
\newcommand{\gsim}{\mathrel{\mathop{\kern 0pt \rlap
  {\raise.2ex\hbox{$>$}}}
  \lower.9ex\hbox{\kern-.190em $\sim$}}}
\newcommand{\alt}{\mathrel{\mathop{\kern 0pt \rlap
  {\raise.2ex\hbox{$<$}}}
  \lower.9ex\hbox{\kern-.190em $\sim$}}}

\begin{document}

\title{Physics at Underground Laboratories: Direct Detection of Dark Matter }

\author{Igor G Irastorza$^*$}

\address{Laboratory of Nuclear and Astroparticle Physics. Universidad de Zaragoza, C/Pebro Cerbuna 12, 50009 Zaragoza, Spain\\
E-mail: Igor.Irastorza@cern.ch \footnote{Although the talk at the
conference included also a brief review of double beta decay
experiments, due to space and time constraints, this written
version is focused only on dark matter.}}


\maketitle

\abstracts{Underground laboratories host two kind of experiments
at the frontier of our knowledge in Particle Physics, Astrophysics
and Cosmology: the direct detection of the Dark Matter of the
Universe and the search for the Neutrinoless Double Beta Decay of
the nuclei. Both experimental quests pose great technical
challenges which are being addressed in different ways by an
important number of groups. Here a updated review of the efforts
being done to detect Dark Matter particles is presented,
emphasizing latest achievements.}

\section{Introduction}

Since first suggested by Zwicky in the 1930s, the existence of an invisible and non-conventional
matter as a dominant part of our Universe has been supported by an ever increasing body of
observational data. The latest precision cosmology measurements
\cite{Spergel:2003cb,Spergel:2006hy,Dunkley:2008ie} further constrain the geometry of the Universe
to be flat ($\Omega \sim 1\pm0.04$), and its composition to be mostly dark energy ($\Omega_\Lambda
= 74 \pm 3\%$) and non-baryonic cold dark matter ($\Omega_{c} \sim 21.4 \pm 3\%$), leaving about
$\sim 4.4\%$ for ordinary baryonic matter. Dark energy is a theoretical concept related to
Einstein's cosmological constant, the nature of which is essentially unknown. Dark matter, on the
contrary, could be composed by elementary particles with relatively known properties, and which
could be searched for by a variety of means. These particles must have mass, be electrically
neutral and interact very weakly with the rest of matter. They must provide a way of being
copiously produced in the early stages of the Universe life, so they fill the above-mentioned
$\sim$23\% of the Universe contents. Neutrinos are the only standard particles fitting in that
scheme, but the hypothesis of neutrinos being the sole component of dark matter fails to reproduce
part of the cosmological observations, in particular the current structure of the universe. The
dark matter problem is therefore solved only by going into models beyond the standard model of
elementary particles, among which two generic categories emerge as the best motivated for the task:
WIMPs and axions.

WIMP is a generic denomination for any Weakly Interacting Massive
Particle. A typical example of WIMP is the lightest supersymmetric
particle (LSP) of SUSY extensions of the standard model, usually
the neutralino. They would have been thermally produced after the
Big Bang, cooled down and then frozen out of equilibrium providing
a relic density\cite{Primack:1988zm,Jungman:1995df}. The
interesting mass window for the WIMPs spans from a few GeV up to
the $\sim$ TeV scale, but can be further constrained for specific
models and considerations.

Axions, on the contrary, are light pseudoscalar particles that are
introduced in extensions of the Standard Model including the
Peccei-Quinn symmetry as a solution to the strong CP
problem\cite{Peccei:1977hh}. This symmetry is spontaneously broken
at some unknown scale $f_a$, and the axion is the associated
pseudo-Goldstone boson \cite{Weinberg:1977ma,Wilczek:1977pj}. The
axion framework provides several ways for them to be produced
copiously in the early stages of the Universe, which makes it a
leading candidate to also solve the dark matter
problem\cite{Raffelt:1990yz,Turner:1989vc}.

The hypothesis of axions or WIMPs composing partially or totally
the missing matter of the Universe is specially appealing because
it comes as an additional bonus to what these particles were
originally thought for, i.e. they are not designed to solve the
dark matter problem, but they may solve it. In addition, the
existence of WIMPs or axions could be at reach of the sensitivity
of current or near future experiments, and this has triggered a
very important experimental activity in the last years. The
detecion of WIMPs by direct means, i. e. aiming at their direct
interaction with terrestrial detectors is being pursued by a large
variety of techniques. All of them, however, share the feature
that they are carried out in underground sites, to reduce
backgrounds induced by cosmic rays. In fact, WIMP experiments are,
together with double beta decay, one of the main pillars of
\emph{underground physics}. In the following pages a review is
given of the current experimental efforts to directly detect these
particles. Axions, on the other hand, are not necessarily searched
in underground labs (although some axion detection techniques do
use underground setups), and because of this they are sometimes
dropped out of dark matter reviews. The fact that the "dark matter
community" is substantially polarized towards WIMPs (probably due
to the boost of underground physics in the last years) should not
be regarded as a justification to ignore or minimize the
importance of the axion as a potential candidate for dark matter.
Much on the contrary, for many the existence of the axion is
better motivated by theory arguments other than being a good dark
matter candidate. Axions are in fact being searched for in
experiments \emph{per se}, without the assumption of them
composing the dark matter, something that does not happen with
WIMPs (excepting supersymmetry searches in accelerators). To
compensate a bit this prejudice, and in spite of this review being
focused on underground physics, we will devote a section on axion
searches at the end of it.

Finally, indirect methods, like those looking for the decay products of WIMPs or axions in
astronomical or cosmic rays observations may also put constraints on the properties of these
particles\cite{Fornengo:2006yy,Hooper:2007vy}, although they suffer from extra degrees of
uncertainties, like the phenomenology driving the accumulation of dark matter particles in
astrophysical bodies and their decay into other particles, and they are left out of the scope of
the present review.

\section{WIMP searches}

If WIMPs compose the missing matter of the universe, and are
present at galactic scales to explain the observed rotation curves
of the galaxies, the space at Earth location is supposed to be
permeated by a flux of these particles characterized by a density
and velocity distribution that depend on the details of the
galactic halo model\cite{Belli:2002yt,Copi:2002hm,Green:2003yh}. A
common estimate \cite{Lewin:1995rx} (although probably not the
best one) gives a local WIMP density of 0.3 GeV/cm$^3$ and a
maxwellian velocity distribution of width $v_{rms}\simeq270$ km/s,
truncated by the galactic escape velocity $v_{esc}\simeq650$ km/s
and shifted by the relative motion of the solar system through the
galactic halo $v_0=230$ km/s.

The direct detection of WIMPs relies on measuring the nuclear recoil produced by their elastic
scattering off target nuclei in underground detectors\cite{Smith:1988kw}. Due to the weakness of
the interaction, the expected signal rates are very low ($1-10^{-6} $ c/kg/day). In addition, the
kinematics of the reaction tells us that the energy transferred to the recoiling nuclei is also
small (keV range), which in ionization and scintillation detectors may be further quenched by the
fact that only a fraction of the recoil energy goes to ionization or scintillation. These generic
properties determine the experimental strategies needed. In general, what makes these searches
uniquely challenging is the combination of the following requirements: thresholds as low as
possible, and at least in the keV range; ultra low backgrounds, which implies the application of
techniques of radiopurity, shielding and event discrimination; target masses as large as possible;
and a high control on the stability of operation over long times, as usually large exposures are
needed.

\begin{figure}[t]
\centering\mbox {\epsfig{file=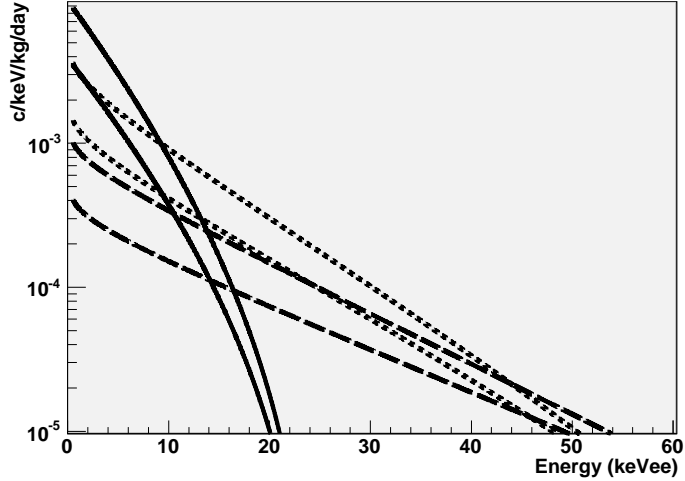,width=10.cm}}
\caption{Example WIMP-induced nuclear recoil spectra (in electron
equivalent energy) for a spin-independent nucleon cross section of
10$^{-42}$ cm$^2$ and both 100 GeV and 500 GeV WIMP mass, for 3
example target materials: Germanium in ionization mode (dots),
Xenon (solid) and Argon (dashed).} \label{WIMPspc}
\end{figure}

Even if these strategies are thoroughly pursued, one extra
important consideration is to be noted. The small WIMP signal
falls in the low-energy region of the spectrum, where the
radioactive and environmental backgrounds accumulate at much
faster rate and with similar spectral shape. Several calculated
WIMP spectra are shown in figure \ref{WIMPspc} for several target
nuclei and some selected input parameters. The pseudo-exponential
shape of these spectra makes WIMP signal and background
practically indistinguishable by looking at their spectral
features. If a clear positive detection is aimed for, then more
sophisticated discrimination techniques and specially more
WIMP-specific signatures are needed. Several positive WIMP
signatures have been proposed, although all of them pose
additional experimental challenges. The first one is the
\emph{annual modulation}\cite{Drukier:1986tm} of the WIMP signal,
reflecting the periodical change of relative WIMP velocity due to
the motion of the Earth around the Sun. The variation is only of a
few \% over the total WIMP signal, so even larger target masses
are needed\cite{Cebrian:1999qk} to be sensitive to it. This signal
may identify a WIMP in the data, provided a very good control of
systematic effects is available, as it is not difficult to imagine
annual cycles in sources of background. A second WIMP signature is
the \emph{A-dependence signature}\cite{Smith:1988kw}, based on the
fact that WIMPs interact differently (in rate as well as in
spectral shape) with different target nuclei. This signature
should be within reach of set-ups composed by sets of detectors of
different target materials, although the technique must face the
very important question of how to assure the background conditions
of all detectors are the same. Finally, the \emph{directionality
signature}\cite{Spergel:1987kx} is based on the possibility of
measuring the nuclear recoil direction, which in galactic
coordinates would be unmistakably distinguished from any
terrestrial background. This option supposes an important
experimental challenge and it is reserved to gaseous detectors,
where the track left by a nuclear recoil, although small, may be
measurable.

\begin{figure}[t]
\centering\mbox {\epsfig{file=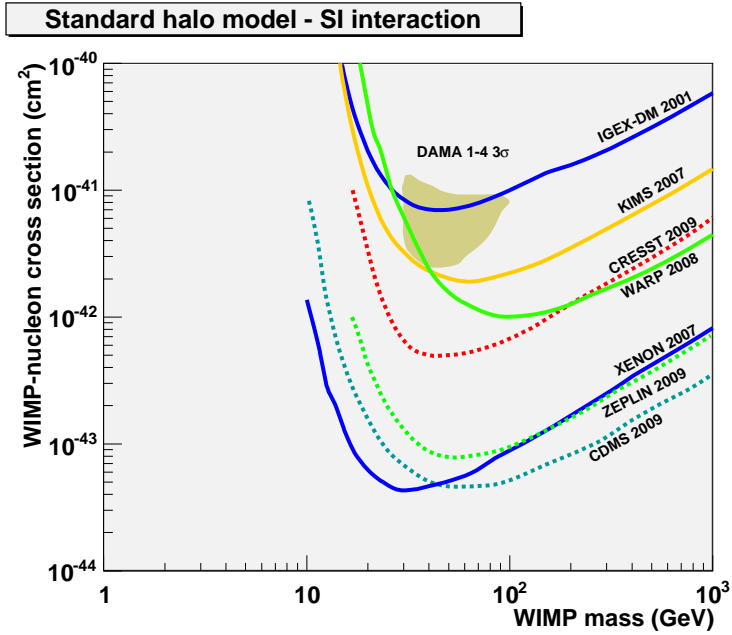,width=10.cm}}
\caption{Exclusion plots for several experiments commented in the
text, as well as the positive region of the DAMA experiment (only
first 4 annual cycles). Standard assumptions for the halo model
and a pure spin-independent WIMP-nucleon interaction have been
considered. See text for more details.} \label{explot}
\end{figure}

Most of the past and current experiments having given the most
competitive results are not sensitive to any of these positive
WIMP signatures, and their reported results are usually exclusion
plots in the ($\sigma_N,M$) plane, obtained by comparing the total
spectra measured directly with the nuclear recoil spectrum
expected for a WIMP (where $\sigma_N$ is WIMP-nucleon cross
section and $M$ the the WIMP mass). These exclusion plots, like
the ones in fig. \ref{explot}, are usually calculated assuming the
standard properties for the halo model previously mentioned, and a
spin independent WIMP-nucleus interaction. But this is an
oversimplification, justified only in part by the need to agree on
some reasonable values for the calculation's unknown inputs for
the sake of comparison between different experiment's
sensitivities. The use of other halo model assumptions, the
introduction of spin-dependent WIMP-nucleus interaction or,
eventually, the invocation of very specific exotic theoretical
frameworks for the WIMP may alter significantly the obtained
exclusion plots\cite{Belli:2002yt,Copi:2002hm,Green:2003yh}. The
study of these effects become especially important in the case of
comparing a positive signal from one experiment with negative
results from others. This is a hot topic in dark matter
conferences, object of much of the discussion regarding the
controversial positive result of the DAMA collaboration (that will
be discussed later). Unfortunately, these problems are inherent to
the model-dependence of exclusion plots and can only be solved by
pursuing experiments capable of getting identifying signals (as
model-independent as possible), like the ones mentioned before.
Fortunately, recent progress in the experimental techniques
promises that experiments in the near future will hopefully have
wider access to these.

In the following section a review of the current status of the
experimental WIMP searches is done. Although some historical notes
are given, it is not aimed at providing an exhaustive and
historical listing of experiments, and it is focussed mainly on
the most relevant detection techniques used with stress in the
latest results and developments. However, the division done in the
first five subsections keep also some chronological dimension as
each of the techniques described has shown their emergence in the
field and sometimes preponderance over the rest for some period of
time in the past. We briefly start with the pioneer
\emph{ionization} detectors (subsection \ref{gedet}), which set a
benchmark in radiopurity and low background techniques, to
continue with the pure \emph{scintillators} (subsection
\ref{scint}) which have up-to-now reached the largest target
masses. \emph{Bolometers} (subsection \ref{bolo}) provided a
breakthrough in the quest for WIMPs in the late 90s by exploiting
very effectively the electron-nuclear recoil discrimination,
although the current lead corresponds to \emph{noble liquid}
techniques (subsection \ref{noble}), which apart from equaling
bolometers in the discrimination capabilities, they seem to
provide better scaling-up prospects. The fifth subsection is
devoted to the gaseous TPCs, a category of experiment that,
although not yet implemented into successful experimental
prototypes is the only one offering access to the WIMP
direccionality signal, and is attiring a renewed attention lately.
The two last subsections are devoted not to different detection
techniques, but to searches specially focused to a specific subset
of WIMP models, like the case of low mass WIMPs, or the
spin-dependent interaction.

Let us finish by saying that for a review containing an historical
listing of the older experiments we refer to
\cite{Morales:2001qq,Morales:2002ud,Morales:2005ky}, which,
although outdated with respect to the latest results, they contain
a very complete and exhaustive listing of experiments up to their
date.

\subsection{The pioneers: ionization detectors}\label{gedet}

The experimental field of dark matter direct detection was born
out of the important advances in background reduction, radiopurity
and shielding techniques exhaustively applied to germanium
detectors in the last 2 decades of the last century, motivated by
the search for the neutrinoless double beta decay of Ge-76.
Nowadays, ionization germanium detectors represent the
conventional approach of a well-known technology, where
radiopurity and shielding techniques have been optimized. In the
last years, their interest for WIMP searches have been reduced in
favor of techniques, discussed in next sections, that exploit
advanced detection features to discriminate between electron and
nuclear recoils, and therefore allowing to go much beyond the best
background levels achieved by Ge detectors. Germanium ionization
detectors remain however a benchmark of the level of raw
background achieved, i.e. only by radiopurity/shielding
techniques, without strong event discrimination techniques, which
may still make them relevant in some specific situations where the
latter are not available, like for example the quest for very low
mass WIMPs discussed in section \ref{lowmass}.

The result released by the IGEX collaboration already 8 years
ago\cite{Morales:2001hj}, shown in fig. \ref{explot} exemplifies
the state-of-the-art in raw background reduction techniques. The
result was obtained with a setup in the Canfranc Underground
Laboratory composed by an ultrapure germanium detector of 2.1 kg,
surrounded by a shielding of ultrapure components, including an
innermost core of 2.5 tons of 2000-year-old archaeological lead
forming a 60 cm side cube, flushed with clean N$_2$ to remove
radon and followed by an extra 20 cm of radiopure lead, a cadmium
sheet, muon vetos and 40 cm of neutron moderator. The achieved
threshold was 4 keV, and the background level was 0.21 c/keV/kg/d
between 4-10 keV (0.10 in 10-20 keV, 0.04 in 20-40 keV), the
lowest raw background level achieved up-to-date in this energy
range.

\subsection{Scintillation detectors}\label{scint}

Scintillation detectors have been extensively used for WIMP
detection, specially because they are the technique which provided
the easiest way to large target masses. It was in fact a setup of
NaI scintillators which first looked for the annual modulation
signature\cite{Sarsa:1996pa}. With no so good prospects concerning
low background capabilities as germanium detectors, scintillation
may however provide a way --although limited-- to discriminate
between nuclear recoils and electron recoils, due to their
slightly different scintillation decay times.

Currently, the DAMA group gathers the expectation of the field
with its claim of observation of an annual modulation signal of
unexplained origin and perfectly compatible with a WIMP of $\sim
52$ GeV and $\sim7.2 \times 10^{-6}$ pb (and standard assumptions
for the halo model). The DAMA experiment in the Gran Sasso
Laboratory, now completed\cite{Bernabei:2003za}, operated 9
radiopure NaI crystals of 9.7 kg each, viewed by two PMTs in
coincidence having gathered 107731 kg day of statistics and
obtaining evidence for the modulation along 7 annual cycles. The
experimental setup was subsequently enlarged up to 250 kg of
target mass in 2005, the so-called DAMA/LIBRA experiment, whose
first results were released in 2007. The DAMA/LIBRA data confirms
the annual modulation seen by DAMA, in intensity, frequency and
phase, and increases its overall statistical significance up to
8.2$\sigma$ CL\cite{Bernabei:2008yi}.

The DAMA positive signal has been ruled out by other experiments
in the standard scenario represented by the exclusion plots of
fig. \ref{explot}. However, in view of the important uncertainties
in the underlying theoretical frameworks and in the galactic halo
models, it is unclear, and a matter of hot discussion, whether all
results are compatible once all uncertainties are taken into
account. It seems that one can always concentrate on a specific
theoretical framework that allows to accommodate both DAMA
positive result and the other exclusion
plots\cite{Bernabei:2003za}. An additional result using the same
target seems to be needed to solve the controversy. An independent
result will come from the ANAIS experiment\cite{Cebrian:2005kz},
currently in the way of instrumenting its $\sim100$ kg of NaI in
the Canfranc Underground Laboratory.

Other promising scintillating material for WIMP searches is CsI,
used by the KIMS experiment\cite{Lee.:2007qn} in the Yangyang
Underground Laboratory in Korea. This material offers a higher
potential of discrimination between nuclear and electron recoils
when compared with NaI, due to the enhanced difference between the
scintillation pulse time pattern of the two kind of events. The
latest KIMS result \cite{Lee.:2007qn} includes 10 kg-y of data
taken with a 36 kg setup, and represent the best exclusion plot by
a pure scintillation setup, excluding WIMP-nucleus cross sections
down to $2\times 10^{-42}$ cm$^{2}$ for WIMP masses around 50-100
GeV, and excluding (always under the standard assumptions) the
DAMA result. This result is especially relevant when compared to
the DAMA positive result, as both experiments share one of the
target nuclei: iodine.

\subsection{Cryogenic detectors}\label{bolo}

Nuclear recoils can also be detected through the heat (phonons)
created in the detector by the recoiling nucleus. This signal is
detectable in bolometers operating at cryogenic temperatures, to
which a suitable thermometer is attached. At those temperatures,
the released heat produces a temperature raise that can be
measurable.

The main advantage of this technique is that most of the energy of
the interaction is visible and therefore no quenching factor must
be applied. Besides, the phonon signal potentially provides the
best energy resolution and thresholds. On the other hand, however,
the operation of cryogenic detectors is a relatively complex
technique facing many challenges when going for larger exposure
times and masses. For the same reason, radiopurity techniques are
also more difficult to apply.

A reference point in pure cryogenics detectors is the pioneering
work of the Milano group, now leading the CUORE/CUORICINO
experiment\cite{Arnaboldi:2002du,Arnaboldi:2004qy} in the Gran
Sasso Laboratory. The CUORE project, designed to search for the
neutrinoless double beta decay of $^{130}$Te, intends the
construction of an array of 988 TeO$_2$ cryogenic crystals,
summing up $\sim 750$ kg of bolometric mass. A first step of the
project, CUORICINO, is already in operation and involves 62
($\sim$40.7 kg) crystals, by far the largest cryogenic mass in
operation underground. Although background levels are still too
high to provide competitive limits in WIMP detection, important
progress is being made and CUORE may have good sensitivity to WIMP
annual modulation\cite{Arnaboldi:2003tu}.

However, cryogenic detectors took the lead in WIMP searches in the beginning of the 2000s because
of the possibility of operating in hybrid mode. Due to the relatively large choice of target
materials available to the cryogenic techniques, and when the material in question is a
semiconductor or a scintillator, the detector could in principle be operated in hybrid mode,
measuring simultaneously the heat and charge or the heat and light respectively. This strategy has
proven to be very competitive and efficient in discriminating nuclear recoils from electron
recoils. Pioneers in this concept are the cryogenic ionization experiments, like CDMS in the Soudan
Underground Laboratory and EDELWEISS\cite{Sanglard:2005we} in the Modane Underground Laboratory.
CDMS took the lead of the race in the standard WIMP exclusion plot since they provided their first
results underground in 2003 \cite{Akerib:2004fq}, with only 19.4 kg-days of effective exposure, and
which went substantially beyond any other exclusion obtained up-to-date. Further improved in 2005
with data from 2 towers\cite{Akerib:2005kh}, it was not challenged until the publication of the
first XENON experiment results (discussed in next section) in 2006. Their most recent published
results\cite{Ahmed:2008eu}, this year, adds data from the full 30-detector setup (19 of them Ge,
and 11 Si, amounting to 4.75 kg of Ge) between October 2006 and July 2007, excluding WIMPs down to
$4.6\times 10^{-44}$ cm$^2$ for a WIMP mass of 60 GeV, and improving slightly beyond the XENON
exclusion limit for WIMP masses above 44 GeV (see Fig. \ref{explot}).

The EDELWEISS collaboration presented similar early results
\cite{Benoit:2001zu,Sanglard:2005we}, but it has not been able to
later improve its sensitivity to the same levels of CDMS mainly
due to the problem of surface events\cite{Fiorucci:2006dx}.
However, the latest advances in their detectors
\cite{edelweiss_taup09} seem to show prospects to reduce these
events in the current physics runs that the new setup, EDELWEISS
II composed by 4 kg of Ge mass (to be extended to 9 kg next year),
is already preforming in the Modane Underground Laboratory.

Although currently less competitive than heat-and-charge, the simultaneous measurement of heat and
\emph{light} has shown very interesting prospects. The ROSEBUD group first applied it
underground\cite{Cebrian:2003jr} (in the Canfranc Lab) and subsequently the CRESST collaboration
presented a competitive exclusion plot obtained with two 300 g CaWO$_4$
prototypes\cite{Angloher:2004tr}. The CRESST setup, installed at Gran Sasso has been improved and
enlarged (CRESST II) to accommodate up to 33 modules of such size working in cryogenic temperatures
(10 mK) and fully shielded. Preliminary results\cite{Angloher:2008jj} from the commissioning run
still with only 2 such modules (total of 600 g of CaWO$_4$) show already a sensitivity down to $4.8
\times 10^{-7}$pb, only one order of magnitude away from CDMS and XENON, as seen in Fig.
\ref{explot}.

A very relevant feature of these results is that tungsten recoils
can be distinguished -with some efficiency- from O or Ca recoils
by virtue of their different ratio heat/light. This improves
substantially the sensitivity of the experiment, as neutrons are
expected to interact more with lighter nuclei, unlike WIMPs. In
addition, recent scintillation studies \cite{Coron:2004iy} have
shown that a large variety of scintillating crystals are
available. This opens the way to use sets of different crystals
operating in this mode to look for the $A$-dependence WIMP
signature. While the use of this signal in conventional detectors
suffers from large uncontrolled systematics derived from the fact
that one cannot assure the background to be the same for different
crystals, light/heat hybrid detectors, sensitive only to neutrons,
may overcome this difficulty. In this line, the ROSEBUD
collaboration has successfully operated underground a set of 3
different bolometers in the same setup \cite{Cebrian:2004ws},
sharing similar external background conditions.

The main challenge in general for hybrid bolometric detectors is
how to cope with the complexity of the technique when going to
larger scales, keeping the ever stronger constraints on
radiopurity and shielding that future setups will have. Present
setups, as mentioned, have achieved target masses at the 1--10 kg
scale. In order to successfully accomplish the next step in scale
($\sim 100$ kg and more) a joint effort is needed. In Europe a new
proto-colaboration has been established,
EURECA\cite{Kraus:2007zz}, which gathers all European groups
mentioned above, EDELWEISS, CRESST and ROSEBUD plus cryogenic
expertise from CERN, as well as new interested groups. EURECA will
work towards a large scale cryogenic facility (up to 1 ton target
mass) that will tentatively be located in the future extension of
the Modane Laboratory, with the goal of solving all related
technical aspects and reaching sensitivities down to $\sim10^{46}$
cm$^2$. In America, the CDMS collaboration have similar plans
(SuperCDMS\cite{Akerib:2006rr}) of evolving towards a ton-scale
facility that would be installed in a deeper location that the
current Soudan site (like SNOLAB for example).

\subsection{Noble liquid experiments}\label{noble}

Experiments using liquid noble gases, especially Argon or Xenon,
should be classified half way between ionization and
scintillation. The scintillation mechanism in noble gases is very
different than in the previous cases, and allows for an improved
discrimination capability by exploiting the different time
patterns of the scintillation pulses of nuclear and electron
recoils or, more efficiently, by using the ratio charge/light when
operating in hybrid mode. This second possibility is available in
two-phase prototypes, where an electric field is applied to
prevent recombination and to drift the electrons to the gaseous
phase where they are detected, either via the secondary
luminescence or by charge amplifiation.

Several groups are developing and using noble liquid detectors for
WIMP searches. They have proven that this technique provides good
prospects of radiopurity and background discrimination and
relatively easy scaling-up.
DAMA/Xe\cite{Bernabei:1998ad,Bernabei:2000qn} is among the
pioneers of the technique, originally motivated for double beta
decay searches. The ZEPLIN collaboration working in the Boulby
Mine Laboratory in UK has also a long-standing program on liquid
Xenon for WIMP searches that has given a series of prototypes of
increasing mass: ZEPLIN I, in pure ionization
mode\cite{Alner:2005pa}, which gave a first competing exclusion
plot in 2004, and ZEPLIN-II and -III working in double phase mode.
But the breakthrough in the field was given by the XENON
collaboration in 2007 when the first 58.6 live days of the XENON10
prototype of 5.4 kg fiducial mass operating in the Gran Sasso
National Laboratory were released. The data excluded WIMP-nucleon
cross secions down to $4.5\times10^{-44}$ cm$^2$ at WIMP masses of
30 GeV, and in general a factor 2 to 10 (depending on the WIMP
mass) better than the best exclusion at the time, that of
CDMS\footnote{as mentioned in the previous section, however, the
latest CDMS result, in 2009, has reached a similar level as that
of XENON10}). The reasons of such good sensitivity are the good
levels of radiopurity achieved the detector, the shieling and
selfshielding, but especially the enhanced discrimination
capability that was achieved even at relatively low energies
(XENON10 was claimed to have significant discrimination down to
4.5 keV nuclear recoil energy). In fact, the behavior of the
ionization of both nuclear and electron recoils in liquid Xenon is
not yet fully understood from fundamental reasons, but, although
this sometimes has been used as a kind of criticism, it is true
that the collaboration has carefully calibrated the effect. The
latest addition of the liquid Xenon scenario are the results
released this same year by the ZEPLIN collaboration, including the
first physics run of the ZEPLIN-III setup\cite{Alner:2007ja},
which has provided an exclusion plot only a factor of 2 to 4 close
to that of XENON10.

The XMASS Japanese collaboration is trying a different approach to
the detection in liquid Xenon. Their prototype is simplified with
respect the previous experiments in the sense that only the
scintillation in the liquid is measured by a set of
photomultipliers that surrounds the Xe mass in all directions (no
double phase operation). The key question beneath this is the
self-shielding concept, consisting in performing fiducial cuts of
the detector to achieve the maximum signal-to-background ratio,
exploiting the fact that external background will interact
primarily in the outer parts of the detector volume. A 100 kg
prototype has been successfully built and
operated\cite{Abe:2008zz} with levels of background in the
fiducial volume compatible with Monte Carlo simulations, although
still not enough to provide competitive exclusion plots. The
collaboration is quickly moving towards the construction of a 800
kg prototype that should be able to achieve much lower backgrounds
in the fiducial volume and get sensitivities\cite{Abe:2008zz} down
to $\sim10^{-45}$ cm$^2$. This detector will be located in the
Kamioka mine. Finally, in US a recent collaboration, LUX, has been
formed with most american groups participating in XENON10 and
ZEPLIN-II experiments in order to build a 350 kg detector using
double phase Xenon technology\cite{luxwebpage}.

At this moment, both germanium bolometers and liquid Xenon
detectors, exemplified by CDMS and XENON respectively, are
providing the best sensitivities for WIMPs, with successful
prototypes at the $\sim5$ kg of fiducial mass scale. Liquid Xenon,
however, have some features that a priori will play a very
important role in the subsequent phases of scaling-up. The
complexity of the liquid Xenon detector does not scale with
detector size as much as in the case of bolometers. One could say
that in the case of bolometers to increase the total size means,
in first approximation, to increase the number of cristal-modules,
and therefore the complexity of the detector (mechanics, readout,
electronics,...) increase linearly with size. In case of liquid
Xenon, the volume of the vessel increases but the readout is
associated to one size of the vessel, and therefore its complexity
scales less than linearly. In addition, the sensitive volume is a
monolithic one in the case of liquid Xenon, and not in the case of
bolometers, which the drawback, for the latter, of the material
needed in between modules and the advantage, for the former, of a
more efficient background reduction strategies based on multi-site
topologies and self-shielding, both concepts of increasing
importance in future bigger detectors. All these arguments are
confirmed in practice by the relatively rapid emergence and
success of the XENON experiment, which has already built the next
prototype XENON100, which will operate between 30 to 50 kg of
fiducial mass, and it is currently being commissioned in the Gran
Sasso Laboratory. In conclusion, short term prospects seem to
favor liquid Xenon technology to lead the quest for WIMPs in the
coming years, at least as long as no irreducible backgrounds (or
signals!) are found.

Apart from liquid Xenon, other noble liquid are being considered. In spite of the current advantage
of Xenon results, the use of liquid Argon or even liquid Neon as proposed by the following
collaborations could emerge as a competitive option with certainly very interesting scaling-up
prospects if the next generation of prototypes show themselves successful. The detection process of
these detectors is similar to that of the liquid Xenon ones, although the discrimination power of
the ratio light/ionizarion seems to be less powerful than in Xenon. In compensation, the scaling-up
prospects, cost and ease of manipulation of Argon surpasses those of Xenon. In Europe two
collaborations are building and testing liquid Argon prototypes for dark matter. Both
collaborations inherit experience in large liquid Argon TPCs from the ICARUS experiment, which
certainly supposes a guarantee of know-how as it was composed of TPC modules of 600 tons of liquid
Argon. The WARP collaboration is for now the only one having produced an exclusion
plot\cite{Benetti:2007cd} which goes down to $10^{-42}$ cm$^2$ at 100 GeV, obtained with a test
chamber of only 2.3 l of liquid Argon accumulating a fiducial exposure of 100 kg-d in the Gran
Sasso Laboratory and also shown in Fig. \ref{explot}. The collaboration is currently commissioning
a second generation prototype of 140 kg in the Gran Sasso\cite{warp_taup}. On the other hand, the
ArDM collaboration has chosen to build a relatively large TPC, of 850 kg of liquid Argon mass,
right from the start. Currently it is being commissioned at surface level\cite{ardm_taup}, at CERN,
and it will be transported underground at a later stage, probably at the Canfranc Underground
Laboratory. Finally, the DEAP/CLEAN collaboration, composed by groups from US and Canada are
building small scale prototypes of both liquid Argon and Neon. The CLEAN
program\cite{McKinsey:2004rk} has a broader scope that includes solar neutrino detection and has
recently proved the feasibility of liquid Neon for such purpose.

\subsection{Directional detectors}

Detectors aiming at measuring the direction of the nuclear recoil
must be put in a different special category. Being extremely
challenging, they have not reached the level of operative
prototypes with any significant sensitivity and remain at an R\&D
phase. However they could have access to the only unmistakable
signature of a WIMP. Many believe that the ultimate identification
of a WIMP as a dark matter component (maybe after its detection in
another non-discrimination experiment) will come from a
directionality experiment. Moreover, the technical advances
achieved in the last years have increased the interest for these
experiments and nowadays a growing community actively explores the
different technological options.

As mentioned in section 1, such signal would suppose a definitive
positive signature of a WIMP and would in addition give
information about how they are distributed in the halo. The
preferred detection medium is gas, where nuclei of $\sim 10-100$
keV could leave tracks in the mm--cm range (depending on the
pressure and nature of the gas), although proposals to scan the
tiny recoils in solid materials (in high resolution nuclear
emulsions) exist\cite{Natsume:2007zz}. The pioneer collaboration
exploring Gaseous Time Projection Chambers (TPCs) for WIMP
directionality is DRIFT\cite{Ayad:2003ph}. DRIFT is developing the
low pressure negative ion TPC
concept\cite{Martoff:2000wi,Ohnuki:2000ex}. Low pressure (40 Torr)
makes the tracks to be relatively long (few cm), and the addition
of electronegative gas (CS$_2$) makes the electrons to be
captured, so the negative ions drift to the avalanche region (a
multiwire proportional chamber) with much smaller diffusion and no
magnetic field is needed. Since 2001, DRIFT has successfully
operated 1 m$^3$ prototypes in the Boulby Mine Laboratory, and
especially with the second generation of prototypes DRIFT-IIa-d
since 2006 has achieved important milestones in terms of
underground operation, background understanding and in particular
the demonstration of sensitivity to the recoil direction sense
(head-tail discrimination)\cite{Burgos:2008jm}.

Despite this progress, sensitivity to nuclear recoil's direction
(especially of low energies, i. e. below 100 keV) is still
challenging, and target masses operated up to now are very small
(few tens of grams), due to the low pressures involved, making
difficult an appropriate strategy towards large scale detector.
Recently, novel readouts based on micropattern technologies are
emerging as an alternative option that could in principle outdo
MWPCs in several aspects, especially in terms of spatial
resolution. In these novel readouts, metallic strips or pads,
precisely printed on plastic supports with photolithography
techniques (much like printed circuit boards), substitute the
traditional wires to receive the drifting charge produced in the
gas. The simplicity, robustness and mechanical precision are much
higher than those of the conventional planes of wires.

Around this basic principle, first introduced by Oed already in
1988\cite{Oed:1988jh}, several different designs have been
developed, which differ in the way the multiplication structure is
done, and that in general are referred to as Micro Pattern Gas
Detectors (MPGD). One of the first clear initiatives to apply this
to WIMP directionality was  the the Japanese NEWAGE collaboration,
which explores since 2003 the use of a MPGD (a microdot structure)
as readout of its first prototype of micro-TPC of $20\times 25
\times 31$ cm$^3$ with low pressure CF4 gas. Since 2007 tests
continue underground in the Kamioka Underground
Observatory\cite{Nishimura:2009zz}, studying aspects like
background, gas purity, scaling-up (bigger prototypes are under
construction) and detector characterization. Up to now, a
threshold of 100 keV has been achieved and an angular resolution
of 55$^\circ$ at 128 torr of pressure, but better numbers are
expected in future prototypes.

Althought probably the most promising MPGD concept for WIMP
searches (and rare event searches in
general\cite{Irastorza_idm06,Dafni_idm08}) is the so-called
Micromesh Gas Structure or Micromegas\cite{Giomataris:1995fq},
created about 14 years ago and actively developed since them by
the CEA/Saclay group led by I. Giomataris. It consists of the use
of a micromesh, suspended over the strip plane by some isolator
pillars, defining a high electric field gap of only 50-100
microns. In this gap, an electron avalanche is produced like the
one in parallel plate chambers, inducing signals in both the mesh
and the strips. Depending on the strip (or pixels) design, the
spatial/topology information of the event can be imaged with
unprecedented precision. Nowadays, the Micromegas concept is
already being used in many particle physics experiments achieving
unique results in terms of temporal, spatial and energy
resolution. In the field of rare events, the use of Micromegas
readouts have been pioneered by the CEA/Saclay and University of
Zaragoza groups, in the CAST
experiment\cite{Zioutas:2004hi,Andriamonje:2007ew,Arik:2008mq} at
CERN. CAST is equipped with a low background Micromegas detector
since 2002\cite{Abbon:2007ug}, and 2 additional ones were
installed in 2007 substituting the former multiwire TPC. CAST has
been a test ground for Micromegas technology, and the continuous
efforts to improve the detectors have given rise to 3 generation
of detector setups (regarding the fabrication method, the detector
materials, their shielding and their electronics and data
reduction treatment), yielding continuously improving background
levels.

Micromegas readouts are being used already by the French MIMAC collaboration, which proposes the
use of a multi-chamber setup composed by microTPCs operating with $^3$He or, alternatively, CF$_4$
at low pressure. The use of light nuclei facilitates the imaging of low energy recoils and the use
of odd-nuclei provides sensitivity of spin-dependent interaction (see next section) and relaxes the
requirement of going to very large masses needed to compete with the experiments searching for
spin-independent interaction. MIMAC has successfully operated a small microTPC module equipped with
Micromegas in $^4$He, has measured ionization quenching factor down to 1 keV, and has recently
obtained first 3D tracks. On the other hand, the DRIFT collaboration has shown interest in the
Micromegas readouts, having successfully tested the negative ion concept with
them\cite{Lightfoot:2007zz}.

More recently, the American collaboration DMTPC has proposed an
optical readout, based on a CCD camera coupled to some appropiate
optics in order to register the scintillation light produced by
the CF$_4$ in the avalanche. They have succeeded in operating a
first prototype and demonstrate 2D imaging of nuclear recoils down
to 100 keV. Future plans include test underground (at WIPP) as
well as development to extract the third dimension.

Although one cannot deny that the construction of a detector with
directional sensitivity at the required mass scale and performance
is still a big challenge, a large and growing development activity
and interest is going on in the last years. This interest is
crystalized in the creation of an international series of
workshops (CYGNUS, Cosmology with Nuclear Recoils\footnote{the
second of which took place at MIT last June 2009}) which gathers
the whole emerging community worldwide. One of the outcomes of the
last gathering has been the preparation of a white paper
presenting the physics case for directional WIMP detectors, the
latest picture of the developments and the justification that the
realization of a directional detector may be realistic in the near
future.

\subsection{Spin-dependent WIMP interactions}

The prejudice of assuming spin-independent (SI) interaction for the WIMP comes because in that case
the interaction for the total nucleus sums coherently over all the nucleons, giving a very
appealing enhancement factor with the mass of the target nuclei $\sim A^2$. This reasoning, used in
all the experiments mentioned up to now, leads to the preferred use of heavy nuclei like Xe of Ge.
However, an important (or even dominant) spin-dependent (SD) component in the WIMP-nucleus
interaction is not at all excluded. In fact, in the case of the WIMP being the lightest
supersymmetric particle (LSP), this is the case for some particular compositions of the neutralino.
In these cases, the WIMP-nucleus is not any more coherent, as the couplings with nucleons of
opposite spins interfere destructively. The target mass does not play any more a special role,
being the nucleus spin (the presence of unpaired neutrons or protons) what matters to determine the
interaction rate, and the sensitivity of the experiment. Experiments sensitive to the standard SD
interaction may be also sensitive to SI models if they contain spin-odd isotopes, like in the case
of CDMS, XENON, KIMS and others. These experiments have published by-products results excluding SD
WIMPs, even if they consider their SI result as the main goal of the
experiment\cite{Akerib:2005za,Angle:2008we}.

However, some experiments are being carried out which are
specifically focused on SD WIMPs. Modest mass of SD sensitive
material can explore neutralino models that are out of reach of
large SD detectors, if the SD interaction is severely suppressed.
This is the argument of experiments like MIMAC, planning to use
light nuclei like $^3$He of CF$_4$ (fluorine). Other experiments
providing primarily SD limits are
PICASSO\cite{BarnabeHeider:2005pg} using superheated C$_4$F$_{10}$
droplets as the active material, or COUPP \cite{Behnke:2008zza}
using the bubble chamber technique with superheated CF$_3$ as
active material.

\subsection{The case for low mass WIMPs}\label{lowmass}

The sensitivity of standard WIMP searches like the ones commented
in the previous pages peaks at WIMP masses around $\sim$100 GeV,
as kinematics favors equal masses of projectile and target. These
masses are also the preferred ones for most popular supersymmetry
models. Lighter neutralinos, however, are not
excluded\cite{Bottino:2005qj}, but the sensitivity of usual
experiments to them drops substantially as can be seen in Fig.
\ref{explot}. Because of this, they have been in fact invoked to
explain the DAMA positive result and reconcile it with all other
limits.

Light ($<$10 GeV) WIMPs would leave very small energy deposits in
the detectors, so sub-keV thresholds are needed to detect them.
Even if such thresholds are achieved, the discrimination
mechanisms that have pushed down the sensitivity of the
experiments during the last decade are not available at such
energies. The search for light WIMPs relies again on pure raw
background reduction, by means of radiopurity and shielding.
Results from "standard" experiments like CRESST or CDMS (but using
data below the discrimination threshold) have been used to extract
limits for low mass WIMPs. However, recently specific setups with
low-threshold detectors have been used to improve limits on light
WIMPs, mainly motivated bye the DAMA signal issue.

The TEXONO experiment\cite{Lin:2007ka} has operated a 4$\times$5 g
Ge detector in a shallow depth site. The smallness of the
detectors allow for thresholds as low as 220 eV, although this
strategy will be difficult to scale-up. More recently, the CoGENT
collaboration has developed a new type of Ge detector, the p-type
point contact (PPC) detector which, profiting from a low-capacity
geometry, allows for both larger masses and low
threshold\cite{Aalseth:2008rx}.

\section{What if there are axions?}

Axion phenomenology\cite{Raffelt:1990yz,Turner:1989vc} depends
mainly on the scale of the PQ symmetry breaking, $f_a$. In fact,
the axion mass is inversely proportional to $f_a$, as well as all
axion couplings. The proportionality constants depend on
particular details of the axion model considered and in general
they can be even zero. An interesting exception is the coupling
axion-photon $g_{a\gamma}$, which arises in every axion model from
the necessary Peccei-Quinn axion-gluon term. This coupling allows
for the conversion of axion into photons in the presence of
(electro)magnetic fields, a process usually called Primakoff
effect and that is beneath all the detection techniques described
in the following.

\subsection{Galactic axions}

Axions could be produced at early stages of the Universe by the
so-called misalignment (or realignment)
effect\cite{Turner:1989vc}. Extra contributions to the relic
density of non-relativistic axions might come from the decay of
primordial topological defects (like axion strings or walls).
There is not a consensus on how much these contributions account
for, so the axion mass window which may give the right amount of
primordial axion density (to solve the dark matter problem) spans
from $10^{-6}$ eV to $10^{-3}$ eV. For higher masses, the axion
production via these channels is normally too low to account for
the missing mass, although its production via standard thermal
process increases. Thermal production yields relativistic axions
(hot dark matter) and is therefore less interesting from the point
of view of solving the dark matter problem, but in principle axion
masses up to $\sim 1$ eV, are not in conflict with cosmological
observations\cite{Hannestad:2005df}.

The best technique to search for low mass axions composing the
galactic dark matter is the microwave cavity originally proposed
in\cite{Sikivie:1983ip}. In a static background magnetic field,
axions will decay into single photons via the Primakoff effect.
The energy of the photons is equal to the rest mass of the axion
with a small contribution from its kinetic energy, hence their
frequency is given by $hf = m_ac^2(1 + O(10^{-6}))$. At the lower
end of the axion mass window of interest, the frequency of the
photons lies in the microwave regime. A high-Q resonant cavity,
tuned to the axion mass serves as high sensitivity detector for
the converted photons.

The Axion Dark Matter Experiment
(ADMX)\cite{Asztalos:2001tf,Asztalos:2003px} has implemented the
concept using a cylindrical cavity of 50 cm in diameter and 1 m
long. The $Q$ is approximately $2\times10^5$ and the resonant
frequency (460 MHz when empty) can be changed by moving a
combination of metal and dielectric rods. The cavity is permeated
by a 8 T magnetic field to trigger the axion-photon conversion,
produced by a superconducting NbTi solenoid.

So far the ADMX experiment has scanned a small axion mass energy,
from 1.9 to 3.3 $\mu$eV\cite{Asztalos:2003px} with a sensitivity
enough to exclude a KSVZ axion, assuming that thermalized axions
compose a major fraction of our galactic halo ($\rho_a=450$
MeV/c$^2$). An independent, high-resolution search channel
operates in parallel to explore the possibility of fine-structure
in the axion signal\cite{Duffy:2005ab}.

Current work focuses on the upgrade of the experimental set-up,
which means basically to reduce the noise temperature of the
amplification stage. This is being done by newly developed SQUID
amplifiers and in a later stage by reducing the temperature of the
cavity from the present 1.5 K down to below 100 mK by using a
dilution refrigerator. These improvements will allow ADMX to
increase the sensitivity to lower axion-photon coupling constants
and also to larger axion masses.

\subsection{Solar axions}

Axions or other hypothetical axion-like particles with a
two-photon interaction can also be produced in the interiors of
stars by Primakoff conversion of the plasma photons. This axion
emission would open new channels of stellar energy drain.
Therefore, energy loss arguments constrain considerable axion
properties in order not to be in conflict with our knowledge of
solar physics or stellar evolution\cite{Raffelt:1999tx}.

In particular, the Sun would offer the strongest source of axions
being a unique opportunity to actually detect these particles. The
solar axion flux can be estimated\cite{vanBibber:1988ge} within
the standard solar model. The expected number of solar axions at
the Earth surface is $\Phi_a=(g_{a\gamma}/10^{-10}\, \rm
GeV^{-1})^2\,3.54\times10^{11}~\rm cm^{-2}~s^{-1}$ and their
energies follow a broad spectral distribution around $\sim$4~keV,
determined by solar physics (Sun's core temperature). Solar
axions, unlike galactic ones, are therefore relativistic
particles.
%

These particles can be converted back into photons in a laboratory
electromagnetic field. Crystalline detectors may provide such
fields \cite{Paschos:1993yf,Creswick:1997pg}, giving rise to very
characteristic Bragg patterns that have been looked for as
byproducts of dark matter underground experiments
\cite{Avignone:1997th,Morales:2001we,Bernabei:2001ny}. However,
the prospects of this technique have been proved to be rather
limited \cite{Cebrian:1998mu}, an do not compete with the
experiments called ''axion helioscopes''
\cite{Sikivie:1983ip,vanBibber:1988ge}, which use magnets to
trigger the axion conversion. This technique was first
experimentally applied in \cite{Lazarus:1992ry} and later on by
the Tokyo helioscope~\cite{Moriyama:1998kd}, which provided the
first limit to solar axions which is "self-consistent", i.e,
compatible with solar physics. Currently, the same basic concept
is being used by the CAST collaboration at CERN
\cite{Zioutas:1998cc,Zioutas:2004hi} with some original additions
that provide a considerable step forward in sensitivity to solar
axions.

The CAST experiment is making use of a decommissioned LHC test
magnet that provides a magnetic field of 9 Tesla along its two
parallel pipes of 2$\times$14.5 cm$^2$ area and 10 m length,
increasing the corresponding axion-photon conversion probability
by a factor 100 with respect to the previous best implementation
of the helioscope concept\cite{Zioutas:2004hi}. The magnet is able
to track the Sun by about 3 hours per day, half in the morning and
half in the evening. At its two ends x-ray detectors are placed,
at the "sunrise" side, a Micromegas detector\cite{Abbon:2007ug}
and a CCD\cite{Kuster:2007ue}, and at the "sunset" side two
additional Micromegas detectors, installed in 2007 replacing the
former TPC\cite{Autiero:2007uf}. All of the detector setups are
conceived following low background techniques (shielding,
radiopure materials). The CCD is coupled to a focusing X-ray
device (X-ray telescope)\cite{Kuster:2007ue} that enhances its
signal-to-background ratio by two orders of magnitude. Both the
CCD and the X-ray telescope are prototypes developed for X-ray
astronomy.

The experiment already released its phase I results form data
taken in 2003 and 2004 with vacuum in the magnet
bores\cite{Zioutas:2004hi,Andriamonje:2007ew}. No signal above
background was observed, implying an upper limit to the
axion-photon coupling $g_{a\gamma} < 8.8 \times 10^{-11}~{\rm
GeV}^{-1}$ at 95\% CL for the low mass (coherence) region $m_a
\alt 0.02~{\rm eV}$. Since 2006 the experiment runs its second
phase, which makes use of a buffer gas inside the magnet bores to
recover the coherence of the conversion for specific axion masses
matching the effective photon mass defined by the buffer gas
density. The pressure of the gas is changed in discrete small
steps to scan the parameter space above $m_a \sim 0.02$ eV. The
$^4$He Run taken in 2006\cite{Arik:2008mq}, allowed to scan axion
masses up to 0.39 eV, for axion-photon couplings down to about
$2.2 \times 10^{-10}~{\rm GeV}^{-1}$, entering into the QCD axion
model band. Due to gas condensation, in order to go to higher
pressures, the experiment switched to $^3$He as buffer gas in
2007. The experiment is currently immersed in the $^3$He Run since
beginning of 2008. It should last until end of 2010 and should
allow us to explore up to 1.2 eV in axion mass approximately,
overlapping with the CMB upper limit on the axion mass discussed
above. At the moment of writing this paper, the experiment has
explored a region of axion masses up to about $m_a \sim 0.70$
eV\cite{cast_taup}. Everyday a new thin slice of untouched
parameter space is being explored. Due to the sharp coherence
effect, and to the fact that the parameter space to which we are
sensitive now is populated by realistic QCD axion models and not
excluded by previous experiments, a clear positive signal in CAST
may appear at any moment.

\section{Conclusions}

A review of the current status of the experimental searches for
WIMPs and axions has been given. The field lives a moment of great
activity, triggered by the fact that very well motivated
theoretical candidates could be within reach of present
technologies. The next years will witness the results of many very
interesting developments currently ongoing to define and operate a
new generation of experiments, well into the region of interest
for both WIMPs and axions.

\section{Acknowledgements}

I would like to dedicate the present review to the memory of prof.
Julio Morales. I am indebted to him for many years of
collaboration and mentoring in the field of low background
techniques. I want to thank the organizers of the International
Meeting on Fundamental Physics 2009 for inviting me to give the
present review and for the friendly atmosphere of the meeting.

\bibliography{igorbib}


\end{document}